# High-Polarization-Extinction Raman Conversion in Gas-Filled Polarization-Maintaining Hollow-Core Fibers

Xianhao Qi,[1,2,†] Pau Arcos,[3,†] Yizhi Sun,[1,2] Shoufei Gao,[1,2,4] Wei Ding,[1,2] David Novoa,[3,5,6,*] and Yingying Wang[1,2,4*]

[1] College of Physics & Optoelectronic Engineering, Jinan University, Guangzhou 510632, China
[2] Guangdong Provincial Key Laboratory of Optical Fiber Sensing and Communication, Institute of Photonics Technology, Jinan University, Guangzhou 510632, China
[3] Department of Communications Engineering, University of the Basque Country (UPV/EHU), Bilbao, 48013, Spain.
[4] Linfiber Technology (Nantong) Co., Ltd., Jiangsu 226010, China
[5] EHU Quantum Center, University of the Basque Country(UPV/EHU), Bilbao, 48013,Spain.
[6] IKERBASQUE, Basque foundation for Science, Bilbao, 48009, Spain.

Email: *david.novoa@ehu.eus; *wangyy@jnu.edu.cn

**Abstract:** Gas-filled hollow-core fibers (HCFs) have emerged as a versatile platform for high-power nonlinear optics, enabling phenomena from ultrafast pulse compression to broadband frequency generation. However, the lack of robust polarization control has remained a critical obstacle to the deployment of gas-based fiber sources. Here, we overcome this bottleneck by demonstrating the generation of highly-polarized Stokes light via stimulated Raman scattering (SRS) in a nitrogen-filled polarization-maintaining anti-resonant hollow-core fiber (PM-HCF). By exploiting the strong structural birefringence of the fiber, the Raman interaction becomes polarization-decoupled along the principal birefringence axes, leading to threshold-selective Raman amplification and an intrinsic polarization purification mechanism. As a result, the vibrational Raman Stokes emission exhibits a polarization extinction ratio (PER) of 35 dB, even when the incident pump PER is as low as ~2 dB. Through analytical theory and numerical modeling, we validate the underlying polarization-selective Raman dynamics and identify the fiber platform as the dominant factor governing the observed PER saturation. We further show that this high polarization purity and high conversion efficiency is maintained under tight bending conditions with radii down to 5 cm, in stark contrast to conventional non-PM-HCF. These results establish PM-HCFs as a robust and scalable architecture for generating polarization-stable, frequency-shifted light, and indicate that polarization may be treated as an actively engineerable degree of freedom in gas photonics, paving the way toward deployment-ready gas-based fiber sources for precision metrology, quantum communication, and coherent sensing.

## 1. Introduction

Hollow-core fibers (HCFs) have been identified as ideal platforms for gas–light interactions—collectively known as "Gas Photonics"[1-4]. When gases, which are ideal nonlinear media with high damage thresholds and broadband transparency, are combined with HCFs that confine optical fields within low-refractive-index cores, the interaction length between light and gas can be extended to meter scales while sustaining multi-gigawatt-level peak powers. This unique configuration enables a series of nonlinear frequency-conversion processes that are inaccessible in conventional solid-core fibers or bulk-optics systems. Representative examples include femtosecond pulse compression down to the single-cycle (<3 fs) regime [5]; Raman frequency conversion extending into the mid-infrared (2 - 5 μm) [6-8]; and vacuum-UV (≈141 nm) [9]; dispersive-wave emission reaching the ultraviolet and vacuum-UV (≈120 nm) [10,11]; Raman frequency-comb or supercontinuum generation spanning multiple octaves [12,13]. These advances demonstrate that gas-filled HCFs open new regimes of wavelength, pulse duration, and optical power that conventional laser architectures cannot reach.

Despite these achievements, a fundamental limitation has persistently constrained the practical deployment of gas-based lasers: the lack of robust polarization control in the generated emission. In conventional gas-filled HCFs, the output light rapidly loses polarization purity during propagation. Random micro-birefringence, bending, and twist perturbations induce coupling between orthogonal polarization modes, degrading the PER and destabilizing the output state of polarization. This polarization instability constitutes a system-level bottleneck that fundamentally limits the utility of gas lasers in precision and deployed environments. Many key applications—such as precision interferometric and metrology [14], fiber-optic gyroscopes [15], coherent communication [16], coherent beam combining [17], and quantum communication [18,19] critically rely on high-PER, polarization-stable light sources. In such systems, polarization noise is directly converted into amplitude noise, phase noise, or measurement bias, undermining both performance and long-term stability.

Overcoming this bottleneck requires an HCF platform that simultaneously offers high damage threshold, broadband low-loss guidance and robust polarization maintenance. This requirement effectively excludes narrow-band photonic-bandgap HCFs, which, although capable of polarization-preserving guidance [20], suffer from low damage threshold and narrow bandwidth. Consequently, research efforts have shifted towards anti-resonant HCFs (AR-HCF). Owing to their wide transmission windows, high resilience to photo-induced damage, good polarization purity [21], and structurally simple geometries, AR-HCFs provide an attractive platform for efficient broadband nonlinear frequency conversion. Achieving polarization maintenance in AR-HCFs, however, remains challenging, as introducing sufficient structural birefringence often comes at the expense of increased loss or reduced bandwidth. Recent breakthroughs—such as the bi-thickness semi-tube design [22,23] or polarization-dependent-loss (PDL) engineered PM-HCF [24]—have demonstrated that strong, stable birefringence can coexist with wideband, low-loss guidance, giving rise to the feasibility of employing PM-HCFs for gas-photonics applications.

Against this, we demonstrate that the introduction of PM-HCFs enables a qualitatively new regime of gas-based nonlinear frequency conversion by restoring and reinforcing polarization control. Using stimulated Raman scattering (SRS) in nitrogen as a proof-of-concept, we reveal that the strong structural birefringence within the fiber effectively preserves the polarization state of the excited vibrational Raman coherence. As a result, the SRS process preferentially affects pump components aligned with the principal birefringence axes while suppressing orthogonal components, resulting in a distinct mechanism of polarization purification. Experimentally, the generated vibrational Stokes emission exhibits a polarization extinction ratio (PER) as high as 35 dB, even when the pump itself is only weakly polarized, with a PER of around 2 dB. This constitutes a fundamentally different polarization paradigm for fiber-based gas Raman lasers, contrasting with the most common scenario where the polarization properties of the frequency down-shifted output are generally inherited, and often degraded, from the pump source. Notably, this observed saturation value at 35 dB is dictated by the intrinsic birefringence limit of the fiber rather than by the nonlinear interaction itself, indicating that further improvements in fiber design could enable substantially higher PER values. Hence, while the PER of the majority of commercial laser sources does not usually exceed ~ 20 dB, our approach might in principle readily provide improved polarization purity at any wavelength from the ultraviolet to the mid-infrared, extremes where high-quality polarizing optics are rather inefficient, not cost-effective and, in some cases, even hardly accessible.

Beyond SRS, our results reveal a new paradigm for gas-based photonics: polarization can be treated as an actively engineerable dimension of nonlinear frequency conversion, on equal footing with wavelength, bandwidth, and efficiency. By shifting polarization control from an external constraint to an inherent property of the gas-light interaction within the fiber, PM-HCFs enable gas-based light sources that are simultaneously spectrally agile, power scalable, and polarization-stable. This capability establishes a practical route towards polarization-stable, broadband, and high-coherence light sources for applications including precision metrology [14], coherent optical sensing [15], and quantum communication [19].

## 2. Theory

In SRS, a strong pump pulse incident upon a molecular sample is inelastically scattered off its constituents and red-shifted to a Stokes band by a frequency shift $\Omega_R$ corresponding to a specific Raman-active molecular transition [25]. At the same time, a coherence wave of synchronized molecular oscillations is generated in the medium, which then enables self-phase-matched coherent amplification of the Stokes signal at the expense of the depletion of the pump light [26]. In general, gases composed of simple homonuclear diatomic molecules such as $N_2$ or $H_2$ exhibit a strongly polarization–dependent Raman response owing to the tensor nature of the third-order nonlinear susceptibility [27]. For instance, SRS in systems where circularly-polarized pump pulses interact with rotational or ro-vibrational degrees of freedom yields amplified Stokes signals whose polarization state is almost orthogonal to that of the impinging pump radiation [28-31]. This observation stems from the fact that Raman gain for molecular transitions involving changes in total angular momentum –such as rotations– is markedly higher when the circular pump and Stokes signals are cross-polarized rather than co-polarized [30]. On the other hand, if the interacting signals are linearly-polarized, both cross- and co-polarized gain scenarios are more balanced [30]. As a result, the Stokes band experiences significant depolarization [32], regardless of the input polarization state of the pump pulses.

The situation is radically different when only pure vibrational transitions (i.e. the fundamental $Q$ branch) are considered. In such scenario, the angular momentum of the excited molecules is conserved during the interaction [33,34], the polarizability of the gas can be regarded as highly isotropic (see Supplementary Material 1) and SRS is therefore reduced to an effective scalar phenomenon [35]. Under these conditions, linearly-polarized pump pulses with an arbitrary polarization orientation just generate co-polarized Stokes light via SRS and the resulting depolarization ratio turns out to be vanishingly small [33].

Thus, let us now consider linearly-polarized pump pulses launched into a PM-HCF. As discussed before, these specialty fibers exhibit relatively high birefringence, which means that the effective refractive indices of the fundamental core modes polarized along the fast ($n_{\hat{f}}$) and slow ($n_{\hat{s}}$) axes are significantly different. As we will discuss in Section 3, in our system this implies that the polarization beat lengths at both pump ($\lambda_P \approx 1064$ nm) and Stokes ($\lambda_S \approx 1415$ nm) wavelengths are much shorter than the fiber samples used in this work (which are several meters long). As a result, any modal field components polarized along the fiber axes are protected along their propagation through a strong reduction of cross-coupling effects that could potentially mix up the polarization states. This greatly simplifies the theoretical description of the system, since a significant energy exchange between orthogonal field components merely based on linear coupling can in principle be disregarded.

Within this framework, we can consider that the electric field of any linearly-polarized signal launched in the PM-HCF can be generally described as $\vec{E} = E_0(\cos\theta \hat{f} + \sin\theta \hat{s})$, with $E_0$ being the amplitude of the field envelope and $\theta$ the relative angle to the fast axis. In other words, both field components will propagate independently along the fiber in the low-power linear regime. In terms of intensities per axis, this implies that $I_P^{\hat{f}} = I_0\cos^2\theta$, $I_P^{\hat{s}} = I_0\sin^2\theta$, where $I_0$ is the total intensity of input pump pulse. Note that this is nothing else but the Malus' law governing the transmission of polarized light through an analyzer.

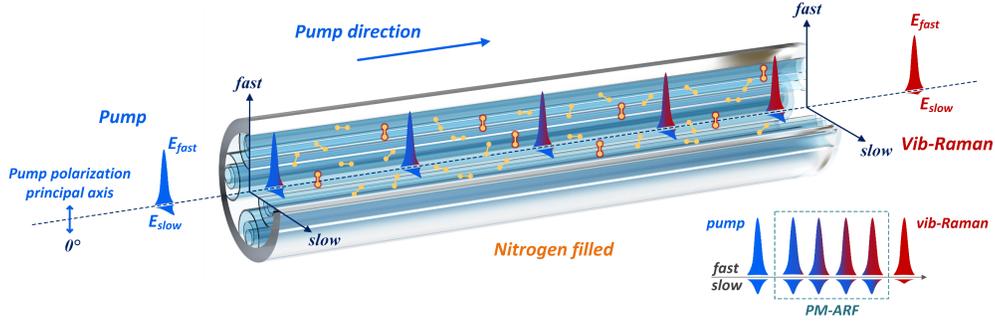

**Fig. 1 Scheme of a coherent pump down conversion via SRS in a gas-filled PM-HCF, where the electric field is mainly aligned with the fast axis.**

Figure 1 schematically illustrates the frequency-conversion process experienced by the incident pump as it propagates through the PM-HCF filled with a Raman-active gas. The polarization direction of the pump is mainly aligned with the fast axis ($\theta \sim 0°$) and a residual part is coupled to the slow axis ($\theta \sim 90°$). Owing to the strong birefringent induced decoupling between both polarization axes and the absence of nonlinear cross-polarizing effects, the frequency conversion proceeds independently along each axis. Consequently, in this example the Raman threshold is first reached along the fast axis, whereas the slow-axis component remains below threshold. Stimulated Raman conversion is therefore predominantly occurring along the fast axis, giving rise to strongly polarization-imbalanced Stokes emission. An equivalent behavior is expected when the pump polarization is aligned with the slow axis, with the roles of the two birefringence axes interchanged.

If the fiber is filled with $N_2$, we can describe the SRS-driven dynamics ocurring inside the gaseous core of the PM-HCF using a two-mode set of coupled Maxwell-Bloch equations[36]:

$$\frac{\partial E_P^l}{\partial z} = -i\kappa_{2,P}\frac{\omega_P}{\omega_S} Q_l E_S^l s_S^l s_P^{l*} - \frac{1}{2}\alpha_P^l E_P^l, \tag{1}$$

$$\frac{\partial E_S^l}{\partial z} = -i\kappa_{2,P} Q_l^* E_P^l s_P^l s_S^{l*} - \frac{1}{2}\alpha_S^l E_S^l, \tag{2}$$

$$\frac{\partial Q_l}{\partial t} + \frac{Q_l}{T_2} = -\frac{i}{4}\kappa_{1,P} E_P^l E_S^{l*} s_P^l s_S^{l*}. \tag{3}$$

Where $Q_l$ and $E_{P,S}^l$ are the amplitudes of the excited molecular coherence and the electric-field envelopes of the Pump (P) and Stokes (S) signals polarized along the direction $l = \hat{f}, \hat{s}$, respectively. The Stokes angular frequency is $\omega_S = \omega_P - \Omega_R$, with $\omega_P$ being the pump angular frequency. As anticipated above and motivated by the experiments, we will restrict our analysis to the fundamental vibrational transition of the nitrogen molecule with frequency $\Omega_R \approx 70$ THz. The spatial phase progression is described by the terms $s_{P,S}^l = \exp(-i\beta_{P,S}^l z)$, where $\beta_{P,S}^l$ are the propagation constants of the pump and Stokes bands, and the $\kappa$ and $\alpha$ are the coupling constants and loss coefficients, respectively (see [36] for more details). We must highlight that, as discussed above, both "fast" and "slow" sets of Eqs. (1-3) are completely independent from each other due to the unique properties of gas-filled PM-HCF.

## 2.1 Analytical solutions for the polarization extinction ratio and the evolution of the Stokes power in the steady-state regime.

To gain some analytical insight into the nonlinear dynamics described by the Maxwell-Bloch Eqs. (1-3), we restrict our study to the asymptotic steady-state SRS regime. This assumption is valid when the pump pulse duration $\tau_P$ is much longer than the molecular dephasing time $T_2$ [36], a requirement that is fulfilled in our system since $\tau_P \sim 6$ ns and $T_2 < 8$ ps. Hence, we can disregard the temporal dynamics of the Raman coherence by setting $\partial_t Q = 0$ in Eq. (3). Additionally, by neglecting the linear attenuation of both pump and Stokes signals $\alpha_{P,S} = 0$ and assuming weak Stokes amplification so that no pump depletion occurs, the evolution of the Stokes field is simply given by the following well-known expression [26]:

$$\frac{\partial E_S^l}{\partial z} = \frac{1}{2} g_P I_P^l E_S^l, \qquad (4)$$

where $g_p$ is the pressure-dependent material Raman gain [36,37] and $I_P^l$ is the pump intensity. Note that in deriving Eq. (4) we have assumed for simplicity that the pressure is constant along the fiber length. However, due to experimental convenience, a positive pressure gradient was implemented to carry out the measurements (see Methods and Supplementary Material), which does not invalidate the conclusions drawn from this simplified model.

The direct integration of Eq. (4) gives the following expression for the exponential amplification of the Stokes intensity along both fiber axes $I_S^l(z) = I_S^l(0)\exp(g_P I_P^l z)$, where $I_S^l(0)$ is the initial seeding for the Stokes emission, which can be originated from spontaneous Raman scattering. In order to be consistent with the experimental results, we always analyze the different observables at the output of the fiber $z = L$, where $L$ is the total fiber length. In this framework, after some straightforward algebra, we arrive at the following expressions for the polarization extinction ratio (PER) of both pump and Stokes fields:

$$PER(Stokes) = \frac{10}{\ln 10} g_P I_P L \cos(2\theta), \qquad (5)$$

$$PER(Pump) = 10 \log_{10}\left(\frac{I_P^{\hat{f}}}{I_P^{\hat{s}}}\right) = 10 \log_{10}(\cot^2 \theta). \qquad (6)$$

Note that Eq. 5 assumes that the strength of the Stokes field is similar in both axes at the fiber input end, something expected in most quantum-fluctuation-initiated SRS dynamics [26,38]. In Fig. 2(a) we display the PER of the pump (yellow-solid line) and Stokes (red-solid line) fields as a function of the pump polarization angle with respect to the fast axis $\theta$. As it can be seen in the graph, for integer multiples of 90° the Stokes PER can be extremely high in absolute value (positive and negative values indicate that either the fast or the slow axes are mainly populated).

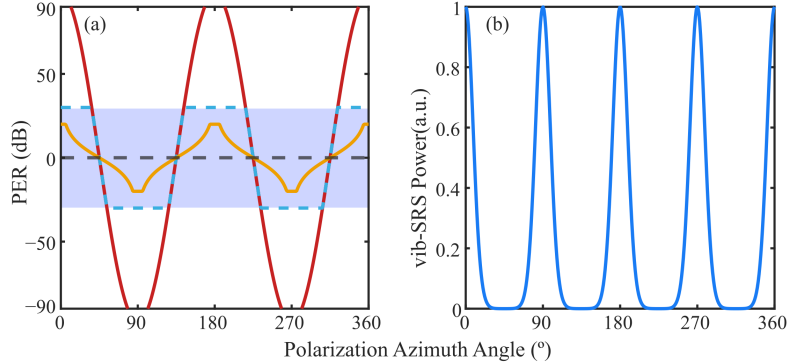

**Fig. 2 Polarization Azimuth Dependence of PER and vib-SRS Power.** (a) Analytical PER of the Stokes (red-solid) and pump (yellow-solid) signals as given by Eqs. (5)-(6) and plotted in decibel scale. The blue-shaded area delimits the dynamic range accessible in the experiments, whereas the blue-dashed line represents the fraction of Stokes PER within that dynamic range. (b) Normalized Stokes power versus polarization angle θ.

In stark contrast, the pump PER is much lower (maximum 17 dB in the experiments) and diverges where the Stokes PER attains its absolute maxima, indicating polarization of the pump along one of the PM-HCF axes. This clearly illustrates one of the main results of this work: In SRS-driven Stokes conversion in gas-filled PM-HCF, the Stokes PER will always be much higher than that of the pump, which might be of interest for the development of novel laser sources featuring high polarization purity.

On the other hand, when comparing with the experimental PER measurements, it is important to understand that their dynamic range is intrinsically limited by several factors such as residual cross-coupling in the fiber, imperfect polarizing optics, etc. These constraint limitations effectively lead to a saturation of the maximum measurable PER in the setup, which is observed at around 35 dB for the Stokes signal (blue-shaded region in Fig. 2a) and causes the actual PER profile to exhibit a series of plateaus (blue-dashed line). Notably, the onset of the PER plateau occurs when the incident pump PER is only 2 dB, highlighting a pronounced band-pass-like behavior of the polarization purification process also observed in the laboratory.

In addition, Fig. 2b shows the dependence of the output Stokes average power $P_S = P_S^{\hat{f}} + P_S^{\hat{s}}$ on the polarization angle $\theta$. The multi-humped curve is normalized to its maximum and represents average power instead of intensity since that is the quantity measured in the laboratory (although they are equivalent). As expected, the Stokes signal peaks for polarization angles that are even multiples of 45°, corresponding to the pump being mainly concentrated along one specific fiber axis, increasing the overall gain. The opposite happens for odd multiples of 45°, where the signal vanishes in this example since the gain is too low to produce any detectable Stokes signal when the initial pump energy is split in half between both polarization axes. However, Stokes conversion might still be observed under these conditions when high gas pressures are used, as we will see in the next section.

## 3. Experiment and simulation

### 3.1 Highly birefringent PM-HCF

To validate the theoretical framework, we conducted experiments using the custom-designed and fabricated PM-HCF shown in Fig. 3a. The introduction of birefringence in such PM-HCF primarily relies on the polarization-dependent anti-crossing effect occurring between the core air modes and the cladding silica modes [39]. This is achieved by designing cladding silica capillaries with varying thicknesses along orthogonal axes. This configuration causes the orthogonally polarized core modes to interact with cladding silica modes possessing different effective refractive indices or coupling strengths, thereby inducing polarization-dependent anti-crossing behavior near specific wavelengths. This results in a divergence in the effective refractive indices of the two orthogonal polarization modes, introducing birefringence [22,23]. This birefringence effectively suppresses random polarization mode coupling caused by external perturbations (e.g., bending, twisting), enabling the fiber to maintain a high PER when excited along its birefringence axes, thus achieving polarization-maintaining transmission. The fiber features an equivalent mode field diameter (MFD) of 15 μm. The transmission spectrum measured using a supercontinuum laser is shown in Fig. 3b, with both the pump (1064 nm) and Stokes (1415 nm) wavelengths falling into the anti-resonant band. Due to the short fiber length, the fiber loss can only be identified using finite-element modelling, showing loss levels of ~3.7 dB/km and ~48 dB/km, respectively. The phase birefringence at the pump and Stokes wavelengths is measured to be $1.48 \times 10^{-5}$ and $8.21 \times 10^{-5}$, respectively, as shown in Fig. 3d. This yields polarization beat lengths on the centimeter scale at both pump and Stokes wavelengths ($B_P \approx 7.2$ cm, $B_S \approx 1.7$ cm), confirming the assumptions made in Section 2.

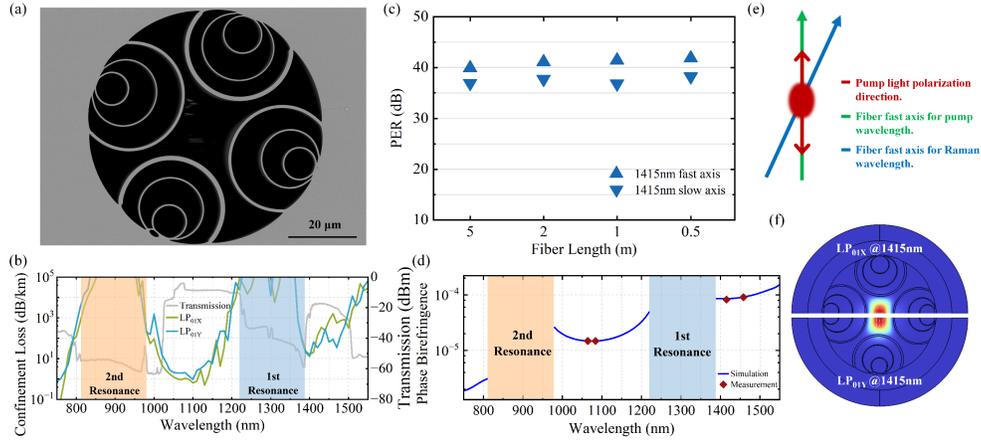

**Fig. 3 Structure and parameters of the PM-ARF.** (a) Cross-sectional scanning electron microscope (SEM) image of the PM-HCF. (b) Transmission spectrum measured using a supercontinuum laser (grey line) and confinement loss derived from FEM simulations (colored lines). The simulated loss is ~3.7 dB/km at the pump wavelength (1064 nm) and ~48 dB/km at the Stokes wavelength (1415 nm). (c) PER measured at 1415 nm (Stokes wavelength) for different fiber lengths. (d) Phase birefringence calculated via simulation (blue curve) and experimental measurement points (red points), showing good agreement between theory and experiment. (e) Orientation of birefringence axes at different wavelengths. (f) Mode fields of polarization modes obtained via FEM simulation.

Using a wavelength-tunable swept-source laser, a PER of 17 dB was measured at 1064 nm for a 10 m-long fiber (see Supplementary Material 2 for details on the measurement procedure). When measuring the PER at the vibrational Stokes wavelength of 1415 nm as a function of fiber length (Fig. 3c), we observed that the Stokes PER increases as the fiber length decreases, eventually reaching stable values of approximately 40 dB. This behavior indicates the presence of polarization cross-talk within the fiber, which accumulates over long propagation distances [40]. Another noticeable feature is that structural asymmetries in the fabricated fiber introduce wavelength-dependent offsets of the principal birefringence axes [15], as shown in Fig. 3e. FEM simulations based on the actual fiber cross-section reveal that the principal axis offset for the $LP01_X$ mode is $\approx 1.189°$ between 1064 nm and 1415 nm, while that of the $LP01_Y$ mode is $\approx 0.942°$ (see Supplementary Material 2.4). Fig. 3f displays the FEM results for the polarization modes at the vibrational Stokes wavelength, where the mode field profile of $LP01_X$ and $LP01_Y$ deviate from a perfect rectangular symmetry, a distortion likewise attributed to the structural asymmetry of the fabricated fiber. The implications of these effects will be discussed in the following section.

### 3.2 PM-SRS dynamics and Stokes light generation

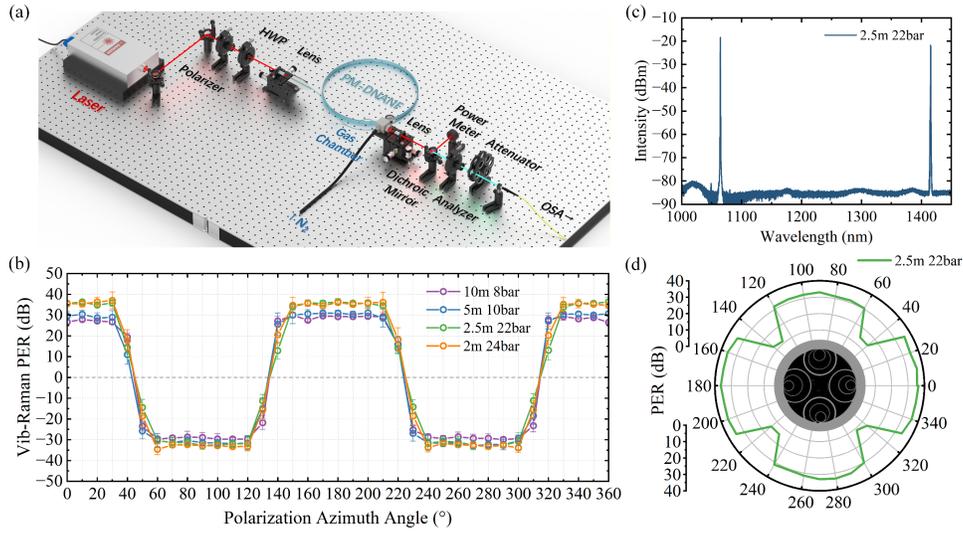

**Fig. 4 Vibrational Raman PER Scan (Measurement setup and results).** (a) Experimental setup. (b) Relationship between polarization azimuth angle and vibrational Stokes PER for different fiber lengths (positive and negative PER values indicate excitation of vibrational Stokes signals along different principal birefringence axes). (c) Spectrum acquired at the output of a 2.5 m-long fiber pumped with light at 0° incidence azimuth. (d) Relationship between vibrational Stokes PER and pump linear polarization azimuth (relative to the fiber end-face) for a 2.5 m-long fiber.

The experimental configuration used to explore SRS in PM-HCF is illustrated in Fig. 4a. 6-ns-long pulses centered at 1064 nm with controllable linear polarization is launched into different lengths of nitrogen-filled PM-HCF to excite SRS-driven dynamics. At the output of the fiber, both average power and PER of the first vibrational Stokes signal at 1415 nm are measured as a function of the input polarization angle (see Methods for details. Specific pressure and pump parameters are detailed in Supplementary Material 3). In Fig. 4b, positive and negative PER values indicate that the polarization of the vibrational Raman signal aligns with different principal axes of the fiber. Experiments across all four lengths demonstrated a significant periodic dependence of the vibrational Raman PER on the azimuth, validating the aforementioned theory. When using fiber lengths of 2 m or shorter, the extremely high gas pressure and pump strength required to excite the signal resulted in relative instability, leading to larger variance in the yellow PER data points in Fig. 4b. Consequently, we selected the configuration of "2.5 m fiber length at 22 bar $N_2$" for detailed analysis, as this configuration yielded relatively stable vibrational Raman signals with high PER. Fig. 4c presents the spectrum collected at a linear polarization incidence azimuth of 0°, where rotational Stokes emission near the pump wavelength is effectively suppressed, resulting in a high-purity energy transfer to the vibrational Stokes signal. Fig. 4d intuitively displays the relationship between the PM-HCF structure and the vibrational Raman PER: even when the pump polarization deviates from the fiber's principal axis by 30 degree, the vibrational Raman PER is maintained at the 35 dB level. As the angle approaches odd multiples of 45°, the excitation direction transitions from the current principal axis to the orthogonal one, causing the PER to tend toward zero and switch signs.

Comparing the PER scanning experiments across different lengths, we found that the maximum achievable Raman PER increased slightly as the fiber length decreased, implying an improvement in polarization purity. At 5 m, the maximum excited vibrational Raman PER was approximately 30 dB; when shortened to 2.5 m, the PER improved to near 35 dB, approaching the fiber's intrinsic PER upper limit at the Raman wavelength (~ 40 dB). We consider this discrepancy acceptable for several reasons. First, due to the asymmetry of the actual fiber structure, the orientation of the principal axes is wavelength-dependent [15], as shown in Fig. 3e. This implies that during the adjustment of the pump linear polarization azimuth, it is impossible to simultaneously align with the principal axes of both the pump and the Stokes waves. Second, during fiber coupling, the mismatch between the PM-HCF mode field shape (Fig. 3f) and the spatial Gaussian beam inevitably results in partial energy coupling into the orthogonal axis; this polarization purity degradation caused by mode field imperfection is unavoidable in experiments. Finally, for current PM-HCFs, there inevitably exist longitudinal inhomogeneities such as surface roughness and micro-bending [41], which lead to distributed crosstalk in the fiber. Under the combined influence of these factors, the vibrational Raman PER (~ 35 dB) remains below the fiber's intrinsic PER limit (~ 40 dB). It is anticipated that under more ideal conditions—such as neglecting the limitations of spatial optical components (e.g., Glan-Thompson prisms) and employing fibers with exceptionally high intrinsic birefringence—the vibrational Raman PER could reach 50 dB or even above 90 dB as theoretically predicted in Fig. 2a.

### 3.3 Simulations

The analytical results presented in Section 2.1 have shed light on the functional behavior of the PER and Stokes power with varying polarization angle of the linearly-polarized pump pulses. However, a direct comparison with the experimental data requires solving Eqs. (1-3) without the approximations made in Section 2.1. Thus, the full model includes the dynamics of the pump pulses and the excited Raman coherence[42], as well as the influence of anti-crossings between core-guided light and modes localized in the semi-tubular cladding [39]. Indeed, these capillary-wall resonances, that cause high loss and uncontrollable dispersion, have been carefully engineered in the fabricated PM-HCFs to place the anti-Stokes signal ($\lambda_{AS} \approx 852$ nm) well within a high-loss band (Fig. 3b), thereby preventing the onset of any anti-Stokes-related effects such as coherent gain suppression [43]. More details can be found in Methods.

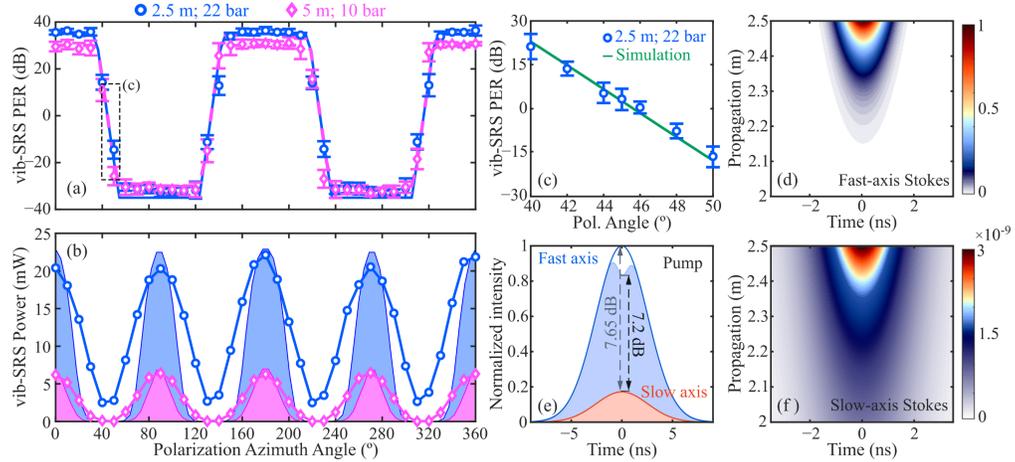

**Fig. 5 Comparison of the numerical simulations.** Solid or dashed line in panel (a) and panel (c) or shaded area in panel (b) with the experimental data (circles or diamonds). Panel (a) and (b) represents the variation of the Stokes PER and power with input angle. Panel (c) provides a close-up of the PER slope between 40° and 50°, and panel (b) represents the evolution of the Stokes power. In blue (pink) is represented a case where we consider 2.5 m (5 m) of fiber and 22 bar (10 bar) of injected pressure. The agreement is very good in panels panel (a), (b) and (c), despite the experimental results in panel (b) displaying a smoother behavior compared to the simulations. We mainly attribute this discrepancy to the Raman gain factor used in the modeling, which may be causing a slight overestimation of the nonlinear dynamics. Panels (d), (e) and (f) show the evolution of the field components for the case with 2.5 m, 22 bar, $\theta = 67.5°$ and pump energies $\varepsilon_{\hat{f}} = 85.35$ μJ, $\varepsilon_{\hat{s}} = 14.65$ μJ. Panel (e) shows the fast/slow-axis pump temporal profiles (lines are input and shaded areas are output), while panels (d) and (f) show the spatio-temporal evolution of the fast/slow-axis Stokes, separately. All panels are normalized to their respective fast-axis signal maximum. Dashed lines in panel (e) indicate pump PER.

The comparison between the experiments and numerical simulations for two different parametric scenarios is displayed in Fig. 5, where panels (a) and (c) show the PER (simulated: blue-solid, pink-dashed line and green-solid; experimental: blue circles and pink diamonds) and panel (b) shows the outcoupled Stokes power for different input angles $\theta$ (simulated: blue-shaded and pink-shaded areas; experimental: blue circles and pink diamonds). The results in blue (pink) correspond to the system displayed in Fig. 4 including a fiber of length $L = 2.5$ m ($L = 5$ m) filled with $p_L = 22$ bar ($p_L = 10$ bar) of $N_2$. As we mentioned before, the simulation results are truncated to approximately ±35 dB for the blue case and ±30 dB for the pink so as to emulate the experimental measuring limitations. Fig. 5c zooms into the region around 45°, where the PER vanishes. The observed negative-slope linear tendency can be predicted by expanding Eq. 5 in a Taylor series around 45°, yielding a $PER \propto -2g_p I_p L(\theta - 45°)$, which indicates that PER changes sign at that angle and is controllable through parameters such as the Raman gain, pump intensity and fiber length (see Supplementary Material). Note that the overall qualitative agreement is excellent in all cases despite having completely different parameters, and the experimental results exhibit all the important features predicted by the analytical theory (Fig. 2). For example, the PER displays a series of plateaus with alternating sign and well-defined transition slopes, while the Stokes power oscillates with the input angle in a controllable way. The convergence of analytical predictions, numerical modelling and experimental results indicate steady-state operation and polarization-decoupled dynamics, which in turn demonstrates how PM-HCF enables unprecedented control over the polarization state of SRS-generated Stokes pulses. For completeness, Fig. 5e, 5d and 5f illustrate the evolution of the field components for the case of 2.5 m, 22 bar, $\theta = 67.5°$ and pump energies $\varepsilon_{\hat{f}} = 85.35$ µJ, $\varepsilon_{\hat{s}} = 14.65$ µJ. Fig. 5e showcases the pump pulse temporal profiles in the fast and slow axes, while the dashed lines denote the PER at each case (around 7 dB), that decreases due to pump-to-Stokes nonlinear conversion along with the preservation of the slow-axis pump below threshold. Fig. 5d and 5f display the spatio-temporal evolution of the Stokes emission along both fast and slow axes. The peak intensity ratio between both projections is about 90 dB, demonstrating the exceptional polarization purity of the Stokes signal with respect to the pump.

### 4. Comparison with Non-PM fiber

To evaluate the intrinsic role of the PM-HCF in the SRS process, we employed a non-polarization-maintaining HCF with a mode field diameter of ~20 µm for comparison. The experimental configuration remained identical to that in Fig. 4a, with the PM-HCF replaced by the non-PM fiber. We selected a 10-meter length of non-PM fiber to excite a sufficiently strong Stokes signal for comparison, while the PM-HCF control group used the most stable "5 m, 10 bar" configuration.

The experimental results are presented in Fig. 6. Fig. 6a displays the cross-sectional structure of the non-PM fiber and the vibrational Raman PER at various azimuths. It can be observed that within the non-PM fiber, the vibrational Raman PER still exhibits a certain periodicity with respect to the polarization azimuth, with a maximum Raman PER of approximately 5.9 dB, far below the 30 dB level achieved with the use of PM fiber. Since non-PM fibers still possess weak birefringence, they exhibit generalized birefringence axes; when the pump azimuth is parallel to these axes, stronger vibrational Raman gain is produced, resulting in relatively high power and PER levels. Furthermore, the PER scanning results are predominantly positive, indicating that this fiber has almost no polarization-maintaining capability along the negative axis direction. The weak periodicity of the vibrational Raman PER in the non-PM fiber stems from its inability to maintain high-birefringence transport for both pump and Stokes signals, which explains the significant degradation in the generated vibrational Raman PER compared to the PM-HCF.

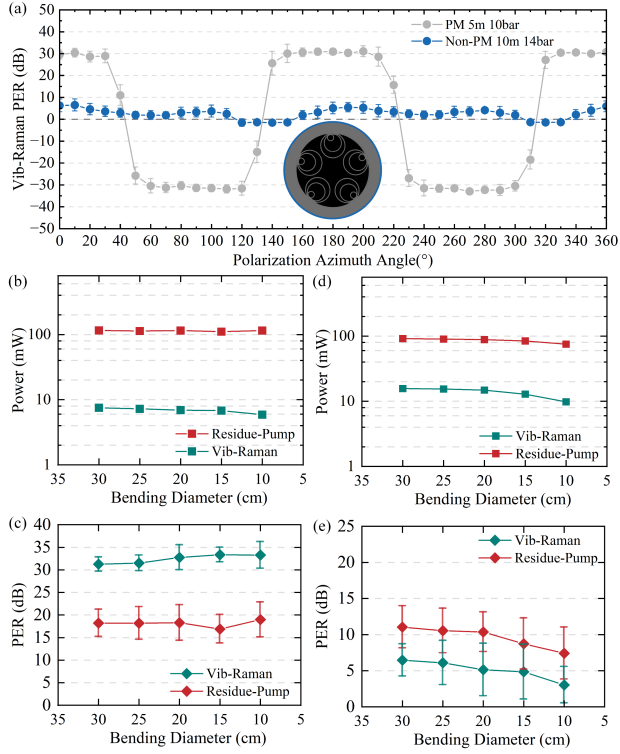

**Fig. 6 PER scanning results for non-PM fiber and bending comparison test.** (a) Cross-section of the non-PM fiber and the variation of the vibrational Raman PER with the linear polarization azimuth angle. (b) Output powers of the PM fiber versus bending diameter. (c) PER evolution of the PM fiber versus bending diameter. (d) Output powers of the non-PM fiber versus bending diameter. (e) PER evolution of the non-PM fiber versus bending diameter.

The superiority of the PM fiber is further highlighted under bending and coiling conditions. In Section 3.2, the fiber coiling diameter was standardized to 30 cm. Here, we sequentially reduced the coiling diameter from 30 cm to 10 cm. The PM-HCF inherently possesses stronger bending resistance. In contrast, the non-PM fiber is highly sensitive to external environmental perturbations, and its transmission capability degrades severely under tight bending. Fig. 6b and 6c illustrate the performance of the PM-HCF under bending perturbations; the power and PER of both the pump and Raman light show no significant changes. This stability is attributed to the fiber's enhanced bending resistance: even at a bending radius of only 5 cm, neither the Raman light intensity nor the PER was affected. Conversely, the non-PM-HCF exhibited distinct differences, as shown in Figs. 6d and 6e. At a bending diameter of 25 cm, both the intensity and PER of the pump and Raman light began to degrade. When the bending diameter reached 10 cm, the pump power decreased by 32.86% compared to the 30 cm case, and the vibrational Raman light decreased by 37.33%, consistent with the non-PM fiber's lack of bending resistance. The degradation in PER was also severe; comparing Fig. 6c and Fig. 6e, the vibrational Raman PER excited in the PM fiber was far greater than the PER of the residual pump, whereas the situation in the non-PM fiber was the reverse. This reinforces the effectiveness of the PM fiber in suppressing and purifying pump polarization noise. Overall, as a platform for SRS, the PM-HCF is significantly more "robust" and better suited for deployment in compact application scenarios.

Using PM-HCFs as the platform for SRS yields vibrational Raman light with a polarization purity that consistently exceeds that of the pump. Furthermore, our additional experiments demonstrated that the PM-HCF-based SRS conversion process can yield high quantum efficiency up to 67.29% along with high PER ~35 dB (see Supplementary Material 4).

Although certain intrinsic factors of the PM-HCF currently restrict the attainable Stokes PER compared to the theoretical estimation, further improvements in the fiber design such as the implementation of high polarization-dependent loss [24] might yield vibrational Raman output with extremely high polarization purity, as predicted in Fig. 2a. This provides a novel approach for stimulated Raman scattering in gas-filled fibers. By optimizing the polarization-maintaining properties of the hollow-core fiber, the polarization output quality of gas Raman lasers [44–46] can be drastically improved. This adds substantial possibilities for the development and practicality of gas-based photonics based on hollow-core fibers.

## 5. Conclusion

In summary, we have demonstrated both experimentally and theoretically that PM-HCFs provide a fundamentally robust solution to the long-standing challenge of polarization instability in gas-based nonlinear optics. Using SRS in a nitrogen-filled PM-HCF, we show that strong structural birefringence enforces polarization-decoupled Raman dynamics, resulting in a vibrational Stokes output with a PER of ~35 dB, substantially exceeding that of the pump. Theoretical analysis suggests that the ultimate polarization extinction ratio can, in principle, exceed 90 dB, indicating significant potential for further improvement through advanced fiber design and system optimization. This level of polarization purity, when compared with the performance of commercial laser sources, whose PER rarely exceeds ~20 dB, together with the demonstrated insensitivity to tight bending down to a radius of 5 cm, demonstrates that polarization-purification approach could provide a practical route to generating highly polarized light across an exceptionally broad spectral range, from the ultraviolet to the mid-infrared, where conventional polarizing optics are often inefficient, prohibitively expensive, or altogether unavailable.

By bridging the gap between high-power gas photonics and the stringent stability requirements of industrial systems, this work expands the research landscape of PM-HCFs in the field of nonlinear SRS conversion and opens new opportunities for polarization-related applications [47-49].

## 6. Methods

### 6.1 Experimental set-up

We employed a microchip laser operating at 1064 nm with a pulse duration of 6 ns, a full-width at half-maximum (FWHM) linewidth of 0.2 nm, and a repetition rate of 1kHz. The single-pulse energy was adjustable up to 300 µJ. The linear polarization angle of the incident light was adjusted using a polarizer (Glan-Thompson prism, Thorlabs, GTH10M) and a zero-order half-wave plate (Thorlabs, WPH0M-1064). In fiber coupling was achieved via a 3-axis translation stage and an aspheric lens. One end of the fiber was sealed within a custom-made gas cell, and the nitrogen pressure inside the fiber was regulated with a pressure reducing valve. A dichroic mirror (Thorlabs, DMLP1180) was used to separate the residual pump light from the vibrational Raman light (1415 nm). The average power of the Stokes signal was measured using a power meter and its PER was characterized using a combination analyzer (Glan-Thompson prism, Thorlabs, GTH10M) and optical spectrum analyzer (Yokogawa, AQ6370D). To investigate the effect of fiber length on the PER plateau region (see Fig. 2a), we performed experiments with fiber lengths sequentially shortened from 10m, 5 m, 2.5 m, to 2 m. Thus, the data points shown in Fig. 4b correspond to the average of the PER measurements at the precise central wavelength of the vibrational Stokes signal (~1415.56 nm), with error bars representing the standard deviation.

### 6.2 Numerical modelling

We numerically solved Eqs. (1−3) using a fourth-order Runge-Kutta scheme with an adaptive evolution step. Apart from the contributions described in Section 3.3, the model also include a positive pressure gradient that appears when the output end of the PM-HCF is pressurized with N$_2$ while the input end is kept under atmospheric pressure. As a result, the pressure distribution along the fiber is no longer homogeneous and obeys the relation $p(z) = \sqrt{p_0^2 - z(p_0^2 - p_L^2)/L}$ [50], where $p_0$ = 1 bar and $p_L$ = 10 bar (or $p_L$ = 22 bar) is the injected gas pressure. Other simulated parameters are very similar to those used in the experiments: 2.5 (or 5) m-long PM-HCF with ~15 μm core diameter and semi-capillary-wall thicknesses of 1.2 μm and 0.875 μm; pump energy $\mathcal{E}$ ~ 100 J, distributed between the two polarization axes as $\mathcal{E}_{\hat{f}} = \mathcal{E}\cos^2\theta$ and $\mathcal{E}_{\hat{s}} = \mathcal{E}\sin^2\theta$, and initial Stokes noise-floor of 600 V/m.

## 7. Acknowledgements

This work was supported by the Scientific Research Innovation Capability Support Project for Young Faculty (ZYGXQNJSKYCXNLZCXM-I18 to Y. W.), National Natural Science Foundation of China (No. 62505111 to Y.S.), the Basic and Applied Basic Research Foundation of Guangdong Province (No. 2025A1515011728 to S. G.), the Guangzhou Science and Technology Program (No. 2024A04J9899 to S. G.), and the Major Key Project of Pengcheng Laboratory (W. D.).

This work was also supported by MICIU/AEI/10.13039/501100011033 and ERDF, EU, under Grant PID2021-123131NA-I00, Grant PRE2022-102843, Grant PID2024-155582OB-C31, and Grant PID2024-158310NB-I00, in part by Gobierno Vasco/Eusko Jaurlaritza under Grant IT1829-26, in part by ELKARTEK SmartμS 2025/00058 and Newhegaz KK 2025/00074, in part by "Translight" Initiative of EHU, and in part by the IKUR Strategy of the Department of Education of the Basque Government under Grant IKUR-IKA-23/03.

## 8. Contributions

X.Q. and P.A. contributed equally to this work. Y.W. and D.N. conceptualized the research. X.Q. performed the PER and Raman generation experiments under Y.W.'s supervision. P.A. and D.N. conducted the simulations and analysis. X.Q. and P.A. drafted the manuscript and figures, with review and editing support from Y.W. and D.N. X.Q. characterized the fibers with help from Y.S. and W.D. S.G. and Y.W. fabricated the PM-HCF and non-PM-HCF. X.Q. analyzed the PER properties of the PM-HCF guided by Y.S.

Supplementary Materials

# High-Polarization-Extinction Raman Conversion in Gas-Filled Polarization-Maintaining Hollow-Core Fibers

Xianhao Qi,[1,2] Pau Arcos,[3,†] Yizhi Sun,[1,2] Shoufei Gao,[1,2,4] Wei Ding,[1,2] David Novoa,[3,5,6,*] and Yingying Wang[1,2,4*]

[1] College of Physics & Optoelectronic Engineering, Jinan University, Guangzhou 510632, China
[2] Guangdong Provincial Key Laboratory of Optical Fiber Sensing and Communication, Institute of Photonics Technology, Jinan University, Guangzhou 510632, China
[3] Department of Communications Engineering, University of the Basque Country (UPV/EHU), Bilbao, 48013, Spain.
[4] Linfiber Technology (Nantong) Co., Ltd., Jiangsu 226010, China
[5] EHU Quantum Center, University of the Basque Country(UPV/EHU), Bilbao, 48013,Spain.
[6] IKERBASQUE, Basque foundation for Science, Bilbao, 48009, Spain.

Email: *david.novoa@ehu.eus; *wangyy@jnu.edu.cn

# Contents

# S1 Theoretical framework

## S1.1 SRS polarization dependence

In this section we will introduce how the polarization of the light affects the Raman scattering. Here we analyze the nonlinear polarization element of third order (the one that gives rise the SRS) to understand the very nature of the vibrational scattering. We represent the electric field $\vec{E}(\vec{r},t)$ vector and the nonlinear polarization $\vec{P}(\vec{r},t)$ of the medium as discrete superpositions of a number of different frequency components, that is:

$$\vec{E}(\vec{r},t) = \sum_m \vec{A}_m(\vec{r})e^{i(\vec{k}\vec{r}+\omega_m t)}, \vec{P}(\vec{r},t) = \sum_m \vec{P}_m(\vec{r})e^{i(\vec{k}\vec{r}+\omega_m t)}, \quad (1)$$

$$E_{-m} = E_m^*; w_{-m} = -w_m, \quad (2)$$

where $m$ represent the different frequency components, $A$ is the spatially slowly varying electric field amplitude and $\vec{k}$ is the wave number vector. The $\omega_m$ represent the different frequency components and

$$P_{i,m} = \sum_{pqr}\sum_{jkl} \chi_{ijnl}(\omega_m;\omega_p,\omega_q,\omega_r)E_{pj}E_{qk}E_{rl} \quad (3)$$

is the electric field dependent part, where $\chi$ is the susceptibility tensor of fourth order.

Here, we are focused in a diatomic molecule ($N_2$) in a gaseous form. For such material, we assume that is isotropic respect to the electric field.

The $\{i,j,k,l\}$ indices represent the different directions of the fields $\{1,2,3\}$ and $\{p,q,r,m\}$ represent the different frequencies involved. For isotropic materials the only tensor components that are independent are those who have even indices (i.e. *ijkl* = 1111, 1122, 1212, 1221). Moreover, $\chi_{1111} = \chi_{1212} + \chi_{1221} + \chi_{1122}$ due to symmetry. The formula stands as follows:

$$P_{1m} = \sum_{pqr} [\chi_{1122}E_{p1}E_{q2}E_{r2} + \chi_{1212}E_{q2}E_{p1}E_{r2} + \chi_{1221}E_{r2}E_{p2}E_{q1}] \quad (4)$$

The sum extends over all available frequency combinations that fulfills $\omega_m = \omega_p + \omega_q + \omega_r$. In our case we will only have Stokes emission so, the allowed frequencies are $\pm\omega_S$ and $\pm\omega_P$. The allowed combinations are $\omega_P = \omega_P + \omega_S - \omega_S$, $\omega_S = \omega_S + \omega_P - \omega_P$, $\omega_P = \omega_P + \omega_P - \omega_P$ and $\omega_S = \omega_S + \omega_S - \omega_S$. If we consider that the susceptibility is independent from any frequency permutation and regroup our terms into vectorial form we get the following expression:

$$\vec{P}_S = 6[\chi_{1122}\vec{E}_S(\vec{E}_P \cdot \vec{E}_P^*) + \chi_{1212}\vec{E}_P(\vec{E}_S \cdot \vec{E}_P^*) + \chi_{1221}\vec{E}_P^*(\vec{E}_S \cdot \vec{E}_P)]$$
$$+3[\chi_{1122}\vec{E}_S(\vec{E}_S \cdot \vec{E}_S^*) + \chi_{1212}\vec{E}_S(\vec{E}_S \cdot \vec{E}_S^*) + \chi_{1221}\vec{E}_S^*(\vec{E}_S \cdot \vec{E}_S)], \quad (5)$$

The terms in the second set of square brackets, which give rise to the nonlinear refractive index but do not contribute to stimulated Raman conversion, can be dropped. It is useful to relate the different susceptibility tensor components with the polarizability anisotropy $\gamma$ as [1,2]:

$$\vec{P}_S = 6\left[\frac{1}{2}(a+c)\vec{E}_S(\vec{E}_P \cdot \vec{E}_P^*) + \frac{1}{2}(a+c)\vec{E}_P(\vec{E}_S \cdot \vec{E}_P^*) + b\vec{E}_P^*(\vec{E}_S \cdot \vec{E}_P)\right], \quad (6)$$

where $a = c = \gamma/9$ and $b = 0$ since, in our system, the tensor is scalar due to the isotropic nature of the vibrational polarizability. Therefore, with this substitution, the nonlinear polarization yields $\vec{P}_S = 2/3\gamma\vec{E}_S(\vec{E}_P \cdot \vec{E}_P^*)$. We can apply a similar process to the pump component of the polarization and get $\vec{P}_P = 2/3\gamma\vec{E}_S(\vec{E}_S \cdot \vec{E}_P^*)$. Now, if we plug it into the wave equation, the evolution of the Stokes field can be described by the following expression:

$$\partial_z\vec{E}_S \propto \vec{E}_S(\vec{E}_P \cdot \vec{E}_P^*) \propto \vec{E}_S|E_P|^2. \quad (7)$$

Assuming no pump depletion and using the Jones formalism, Eq. (7) becomes polarization independent. Such approximation is backed up with experimental data which shows that the Stokes polarization is never emitted cross-polarized respect to the pump polarization [3], i.e. the Stokes is always co-polarized in vibrational transitions of the Q(1) branch [4].

**1.2 Linear approximation of PER around 0 dB**

In Section 2 of the main text, we obtained the polarization extinction ratio (PER) for Stokes emission as:

$$\text{PER}(Stokes) = \frac{10}{\ln 10} g_P I_P L \cos(2\theta), \tag{8}$$

Its cosinusoidal dependence implies near linear behavior around PER = 0 dB, so if we expand around $\theta_0 = 45°$ we get:

$$\cos(2\theta) \approx -2(\theta - 45°), \tag{9}$$

$$\text{PER} \propto -2 g_P I_P L (\theta - 45°). \tag{10}$$

Eq. (10) shows that, in the vicinity of $\theta_0 = 45°$, the PER slope scales linearly with $g_P$, $I_P$, and $L$. As an illustrative example, Fig. S1 confirms that the PER slope gets steeper when we increase the fiber length. This behavior is crucial since these parameters (fiber length, Raman gain, and pump intensity) can be experimentally controlled, enabling access to on-demand Stokes PER and, consequently, enhanced polarization purification.

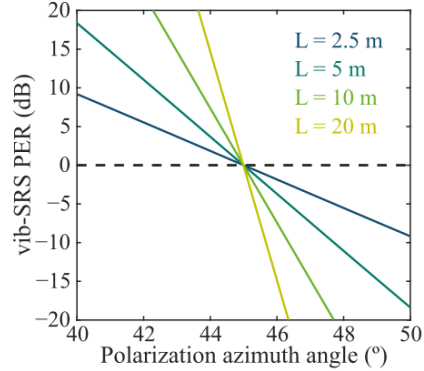

**Fig. S1 shows an illustration of the PER variation with the azimuth angle for different fiber lengths.**

## 2. Experimental and Data Processing Details

### S2.1 Measurement Procedure for Fiber PER Upper Limit

The Polarization Extinction Ratio (PER) is conventionally used to characterize the purity of linear polarization. It is defined as:

$$\text{PER} = 10 \log_{10}\left(\frac{P_1}{P_2}\right). \tag{11}$$

Where $P_1$ and $P_2$ represent the optical power levels along two orthogonal directions. In this work, we measured the PER using an optical spectrum analyzer (Yokogawa, AQ6370D). The measurement setup is illustrated in Fig. S2. A wavelength-tunable swept laser was employed as the light source. To select the polarization of the free-space light, we utilized a polarizer and an analyzer (Thorlabs, GTH10M). By selecting orthogonal directions, the intensity levels along the different birefringence axes of the fiber were measured. Subtracting the spectrum obtained with the polarizer at 90° from that at 0° yields the PER level of the hollow-core fiber at the laser wavelength.

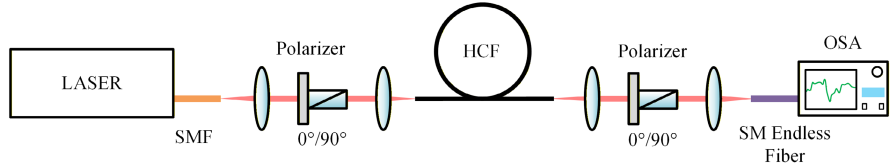

**Figure S2: PER measurement process for the hollow-core fiber.**

Due to the presence of distributed crosstalk, the measured PER of a Polarization-Maintaining Hollow-Core Fiber (PM-HCF) increases as the fiber length decreases. To determine the upper limit of the PER for the PM-HCF used in our experiments within the vibrational Raman band, we sequentially cut back the fiber to lengths of 5 m, 2 m, 1 m, and 0.5 m for measurement. It is important to note that after each cutback, the optical path required realignment to ensure consistent coupling efficiency.

### S2.2 Method for Measuring Vibrational Raman PER

Unlike the setup in Fig. S2, the measurement of vibrational Raman light must be conducted during the Raman generation process. As shown in Fig. S3, an analyzer was inserted after the dichroic mirror (DM). When the half-wave plate was adjusted to the desired angle (during PER scanning experiments), the analyzer was rotated to locate the spectral lines with the strongest and weakest intensities on the spectrometer. By fine-tuning the analyzer, the angular settings corresponding to the maximum and minimum intensities were identified; these angles correspond to the fast and slow axes of the fiber output face. We collected the vibrational Raman signals output from the fiber's fast and slow axes using the spectrometer. The PER data for vibrational Raman scattering were obtained from the spectra collected at the outputs of the fast and slow axes of the fiber, centered at 1415.55 nm. The average values of these PER measurements correspond to the data points presented in Section 3.2, with error bars representing the standard deviation of the PER. All subsequent PER data related to vibrational Raman scattering were acquired using the same methodology.

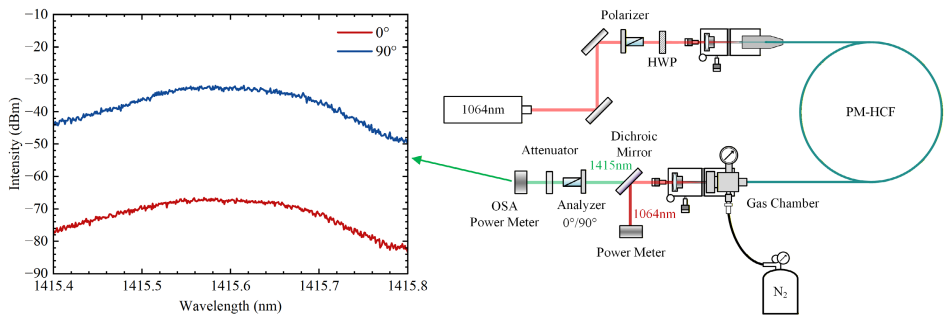

**Fig. S3: Experimental setup for the PER scan measurements.**

Following multiple sets of vibrational Raman PER measurement experiments, we observed that the angular difference between the fast and slow axes of the PM-HCF used in this work was not exactly 90°. This deviation corresponds to the imperfections in the actual fiber structure, indicating that the fiber's birefringence axes are not perfectly collinear and orthogonal.

### S2.3 Operational Details for Gas Filling Experiments

Due to the varying cross-sectional areas of the microstructured regions within the hollow-core fiber, gas flow rates differ across these regions during filling. This discrepancy leads to pressure differentials between regions, as illustrated by the varying shades of blue at different depths in Fig. S4a. Consequently, it is necessary to equilibrate the air pressure among the various regions to prevent structural collapse of the thin microstructured walls caused by pressure gradients.

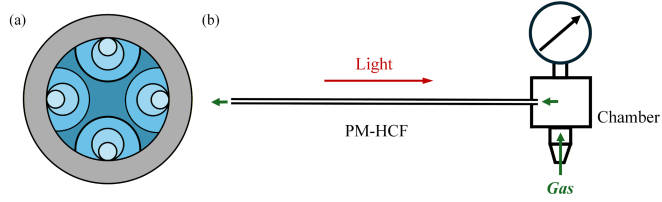

**Fig. S4: Fiber End-Face and Gas Filling Process.** a. Pressure gradient within the fiber. b. Process of filling the gas cell and fiber.

To reduce back-reflection, improve fiber coupling efficiency, and facilitate fiber cut-back for varied measurements, this work employed a single gas cell, creating a semi-sealed gas environment, as shown in Fig. S3b. A pressure gradient exists along the fiber, with higher pressure near the gas cell and lower pressure further away. We estimated the flow rate of gas molecules in different regions based on the approximation of the Poiseuille equation [5] within a continuous medium as:

$$Q = \frac{\pi r^4}{8\eta L}\Delta P. \tag{12}$$

The volume of a capillary section is given by:

$$V = \pi r^2 L. \tag{13}$$

The time required for pressure equilibration in a region can be estimated from $t = V/Q$.

Based on calculations, to balance the pressure difference across different regions of this specific fiber, a waiting time of no less than 20 minutes was required after each pressure increase not exceeding 3 bar. In this work, during gas filling, the pressure was increased stepwise to 2 bar, 4 bar, 6 bar..., waiting for 30 minutes after each increment.

### S2.4 Wavelength-Dependent Fiber Principal Axis (Simulation)

For PM-HCF, the actual fiber structure is not perfectly symmetric, causing the birefringence principal axes to exhibit an offset angle. This offset angle is wavelength-dependent. To investigate the orientation of the principal axes of the PM-HCF used in our experiments at the pump wavelength and the vibrational Raman wavelength, we performed finite element simulations of the actual fiber structure using COMSOL software. By calculating the modes at the precise pump wavelength (1065.45 nm) and the precise vibrational Raman wavelength (1415.55 nm), we obtained the principal axis orientation relationship shown in Fig. S5.

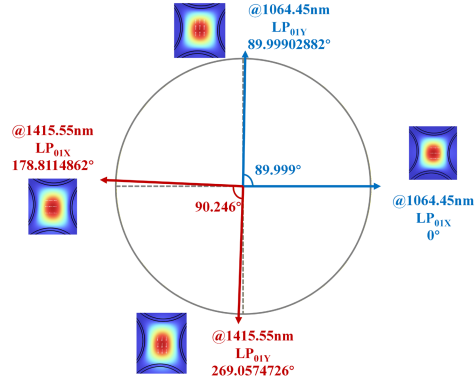

**Fig. S5: Fiber principal axes calculated via finite element simulation.**

We calibrated the principal axis angles for the pump and vibrational Raman waves using the polarization electric field direction of the $LP_{01X}$ mode at the pump wavelength as the 0° reference. The calculations reveal that neither the polarization principal axis of the pump nor that of the vibrational Raman wave is perfectly at 90°, which aligns with the phenomena observed in our experiments. According to the simulation results, the $LP_{01X}$ polarization principal axis at 1064 nm and 1415 nm has a relative offset of 1.189°. For the $LP_{01Y}$ mode, the relative offset is 0.942°.

## S3 Supplementary Data for Vibrational Raman PER Scanning Experiments

### 3.1 Pressure and Pump Parameters for Fibers of Different Lengths

For the vibrational Raman PER experiments with different fiber lengths (Section 3.2 of the main text), we employed different gas pressures and pump intensities. This was necessary to excite sufficient vibrational Raman signals for characterization. The gas pressures and pump intensities used for different fiber lengths are listed in Supplementary Table 1.

**Supplement Table 1. Experimental parameters and Vibrational Raman output.**

| Fiber length(m) | Gas pressure(bar) | Effective pump power (mW) | Vibrational Raman power at 0° (mW) |
|---|---|---|---|
| 10 | 8 | 102 | 7.79 |
| 5 | 10 | 104 | 6.45 |
| 2.5 | 22 | 129 | 19.36 |
| 2 | 24 | 105 | 4.64 |

### 3.2 Comparison with Non-PM Fibers

When conducting comparative SRS excitation experiments using non-PM fibers (Section 4 of the main text), we characterized the vibrational Raman PER and power output, as shown in Fig. S6.

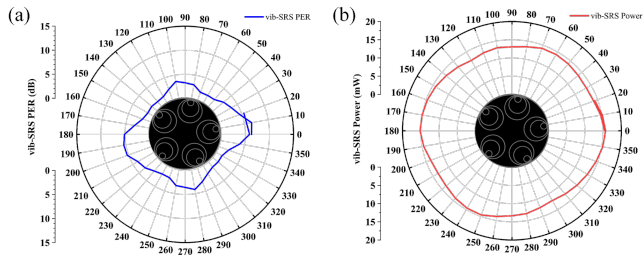

**Fig S6: SRS experiments using non-PM fiber for linear polarization azimuth scanning.** a. Relationship between vibrational Raman PER and incident polarization azimuth. b. Relationship between vibrational Raman power and incident polarization azimuth.

PER scanning was performed by varying the linear polarization azimuth. The variations of vibrational Raman PER and power with the polarization azimuth are presented in Fig. S6. In non-PM fibers, the vibrational Raman signal still exhibits some regular periodicity with the polarization azimuth; the Raman PER excited at the point of highest power is typically the strongest (5.9 dB), while the absolute value of the Raman PER at the lowest power point is closest to 0 dB. Furthermore, the fluctuation of vibrational Raman power with the pump azimuth is minimal, ranging between 12 and 15 mW.

The periodicity of the Raman light in non-PM fiber with respect to the pump azimuth arises because non-PM fibers still possess a weak degree of birefringence. Consequently, generalized birefringence axes exist. When the pump azimuth is parallel to these birefringence axes, stronger vibrational Raman gain is generated, resulting in higher power and PER levels for the excited vibrational Raman light. However, compared to the intense and significant PER periodicity observed in PM fibers, the excitation efficiency and PER purity in non-PM fibers are significantly inferior.

## S4 Optimization of SRS Experiments Using PM-ARF

To explore the achievable Quantum Efficiency (QE) for SRS using the PM-ARF, we optimized the fiber length, gas pressure, and pump power.

### 4.1 Fiber Length

In the initial phase of this work, we conducted SRS experiments using a 10 m fiber. Due to the excessive fiber length, rotational Raman scattering was inevitably excited, accompanied by severe cascading phenomena. The spectrum is shown in Fig. S7a, where rotational Raman cascade lines pumped by the vibrational Raman signal appeared in the 1400 nm–1450 nm range.

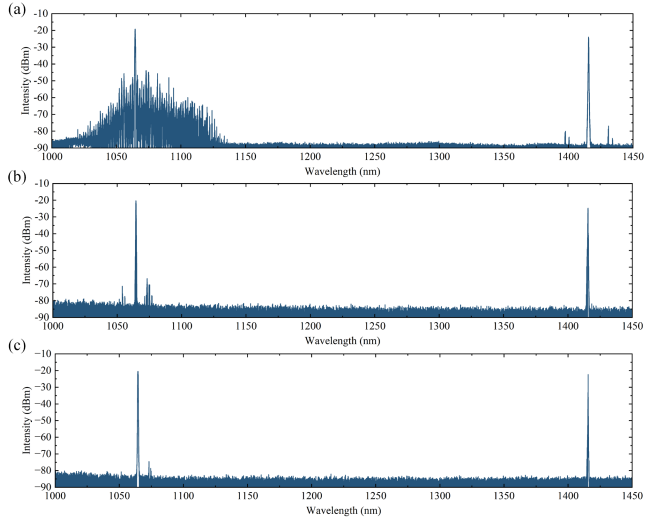

**Fig. S7: Spectra from Gas SRS Experiments Using Fibers of Different Lengths.** a. Excitation spectrum with 10-meter fiber. b. Excitation spectrum with 6-meter fiber. c. Excitation spectrum with 5-meter fiber.

Following multiple optimizations of the fiber length, we found that a 5 m length yielded relatively pure vibrational Raman light. Only minimal rotational lines appeared near the pump (~80 dB), which were negligible. To filter out the effects of interference between rotational and vibrational lines and to ensure efficient Raman conversion, the high-conversion-efficiency experiments were conducted at a length of 5 meters.

### S4.2 Gas Pressure

Using the 5 m PM fiber with an effective average pump power of 141 mW, we increased the gas pressure and recorded the power and PER at different pressures. As shown in Fig. S8a, the vibrational Raman power increased with pressure, accompanied by a decrease in residual pump power. During the pressurization process, the vibrational Raman PER was maintained at 30-35 dB, while the pump PER decreased to approximately 15 dB as the pressure increased. At 34 bar, cascaded rotational Raman scattering pumped by the vibrational Raman signal began to appear (Fig. S8b). Consequently, the pressure was not increased further. At this point, the QE was 66.40%.

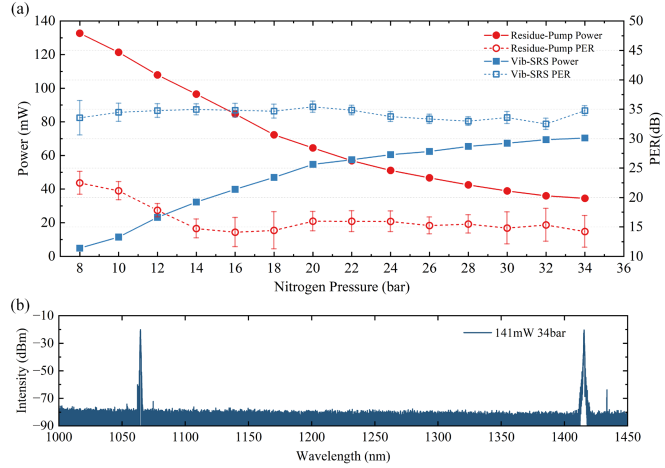

**Fig. S8: Raman Conversion versus Gas Pressure.** a. Vibrational Raman conversion versus increasing gas pressure. b. Output spectrum of the fiber at 34 bar.

### S4.3 Pump Power

Maintaining the gas chamber pressure at 34 bar, we varied the laser power (the microchip laser used in this work requires adjusting the input current intensity to change the laser power). The effective pump power was increased from 100 mW to 170 mW, exciting vibrational Raman light of varying intensities. Fig. S9a displays the excitation under different pump intensities. Notably, an effective pump power of 141 mW corresponds to 7700 mA; when the pump power exceeded 141 mW, cascaded rotational Raman scattering emerged.

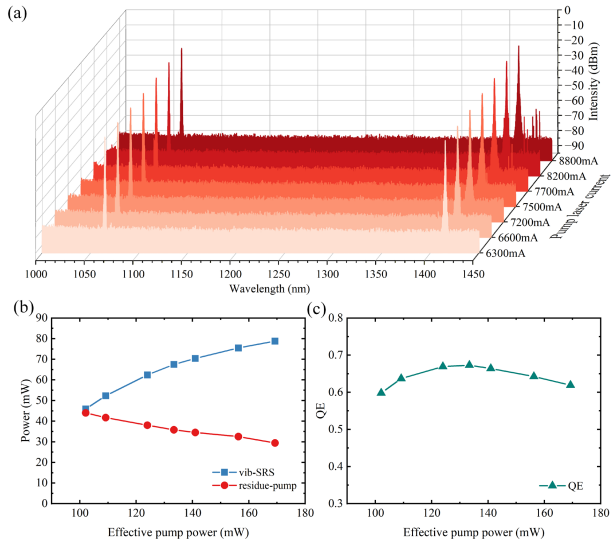

**Fig. S9: Raman Conversion versus Pump Power.** a. Spectra at different pump power levels. b. Vibrational Raman and residual pump power versus pump power. c. Vibrational Raman Quantum Efficiency versus pump power.

Supplementary Table S2 illustrates the relationship between pump current and effective pump power. Combined with Fig. S9b, the highest vibrational Raman power (average power of 78.8 mW) was achieved at a pump intensity of 169.3 mW. However, the highest vibrational Raman quantum efficiency (67.29%) was observed at 133.4 mW. As the pump intensity increased further, a portion of the energy was transferred to the cascaded rotational Raman modes. Thus, indiscriminately increasing the pump intensity eventually leads to a reduction in quantum efficiency.

**Supplement Table 1. Pump current and vibrational Raman conversion performance.**

| Laser current(mA) | Effective pump power (mW) | Vib-SRS power (mW) | Residue-pump power (mW) |
| --- | --- | --- | --- |
| 6300 | 102.1 | 45.9 | 44.0 |
| 6600 | 109.2 | 52.3 | 41.7 |
| 7200 | 124.0 | 62.4 | 38 |
| 7500 | 133.4 | 67.5 | 35.8 |
| 7700 | 141.0 | 70.4 | 34.5 |
| 8200 | 156.3 | 75.5 | 32.5 |
| 8800 | 169.3 | 78.8 | 29.4 |

Through this series of experiments, we demonstrated frequency conversion achieving both high quantum efficiency (max QE = 67.29%) and high linear polarization purity (PER ~35 dB).